# Evidence for geologically recent explosive volcanism in Elysium Planitia, Mars

David G. Horvath[1], Pranabendu Moitra[1], Christopher W. Hamilton[1], Robert A. Craddock[2], Jeffrey C. Andrews-Hanna[1]

[1]Lunar and Planetary Laboratory, University of Arizona, Tucson, Arizona, USA
[2]Center for Earth and Planetary Studies, National Air and Space Museum, Smithsonian Institution, Washington, DC, USA

Corresponding author: David G. Horvath (dhorvath@lpl.arizona.edu)

*Submitted to Icarus*

**Abstract**

Volcanic activity on Mars peaked during the Noachian and Hesperian periods but has continued since then in isolated locales. Elysium Planitia hosts numerous young, fissure-fed flood lavas with ages ranging from approximately 500 to 2.5 million years (Ma). We present evidence for what may be the youngest volcanic deposit yet documented on Mars: a low albedo, high thermal inertia, high-calcium pyroxene-rich deposit distributed symmetrically around a segment of the Cerberus Fossae fissure system in Elysium Planitia. This deposit is similar to features interpreted as pyroclastic deposits on the Moon and Mercury. However, unlike previously documented lava flows in Elysium Planitia, this feature is morphologically consistent with a fissure-fed pyroclastic deposit, mantling the surrounding lava flows with a thickness on the order of tens of cm over most of the deposit and a volume of $1.1–2.8 \times 10^7$ m$^3$. Thickness and volume estimates are consistent with tephra fall deposits on Earth. Stratigraphic relationships indicate a relative age younger than the surrounding volcanic plains and the Zunil impact crater (~0.1–1 Ma), with crater counting suggesting an absolute model age of 53 to 210 ka. This young age implies that if this deposit is of volcanic origin then the Cerberus Fossae region may not be extinct and Mars may still be volcanically active today. This interpretation is consistent with the identification of seismicity in this region by the Interior Exploration using Seismic Investigations, Geodesy, and Heat Transport (InSight) lander, and has additional implications for astrobiology and the source of transient atmospheric methane.

**1. Introduction**

Effusive volcanism dominates the geologic record of Mars (Greeley & Spudis, 1981), from Hesperian-aged (~3.8–3.0 Ga) volcanic plains to Amazonian-aged (<3.0 Ga) flood basalt lava flows of Amazonis and Elysium planitiae. In contrast, evidence of explosive volcanism on Mars is comparatively less common. Although some paterae within Arabia Terra have been attributed to ancient (Noachian- to Hesperian-age) caldera-forming eruptions with associated pyroclastic flows (Michalski & Bleacher, 2013), the best lines of evidence for the occurrence of explosive volcanism on Mars comes from the widespread distribution of clastic material within the Medusa Fossae Formation (MFF). This unit has been interpreted as the remains of a much larger Hesperian tephra deposit, which has now been extensively eroded (Scott & Tanaka, 1986; Greeley & Guest, 1987; Bradley et al., 2002). A pyroclastic origin is generally accepted for the MFF based on its friable nature and the manner in which this material mantles topography unconformably (Bradley et al., 2002), though the source of this material is debated. Multiple provenances have been proposed, including the Tharsis Montes (Hynek et al., 2003; Mouginis-Mark & Zimbelman, 2020), Apollinaris Mons (Kerber et al., 2011a), and pyroclastic material associated with dominantly effusive eruptions sourced from the Cerberus Fossae in Elysium Planitia (Keszthelyi et al., 2000). Apollinaris Patera has also been interpreted as having large Hesperian-aged large pyroclastic flow deposits, owing to the lack of primary lava flow features and deeply eroded, friable nature (Crown & Greeley, 1993; Robinson et al., 1993; Gregg et al., 1996). Similar ash flows may have also originated from the Tharsis Montes (Scott & Tanaka, 1982).

Additionally, a number of smaller Hesperian and Amazonian features have been interpreted as possible pyroclastic eruption products, many of which are contained in the Tharsis region and around Olympus Mons. These include potential spatter deposits, which are similar to those produced by lava-fountaining during effusive eruptions on Earth (Wilson et al., 2009), fine-grained material on Arsia Mons (Mouginis-Mark, 2002) and potential phreatomagmatic pyroclastic cones north of Olympus Mons (Wilson & Mouginis-Mark, 2003a). However, while evidence for relatively pristine pyroclastic deposits exists on the Moon and Mercury (Gaddis et al., 1985; Head et al., 2009), similar pristine deposits have not been documented on Mars despite martian conditions favoring such eruptions and evidence for abundant water in the deep interior and shallow subsurface to help drive the explosive activity (Wilson & Head, 1983; Fagents & Wilson, 1996).

Elysium Planitia, in particular, contains abundant evidence for the interplay between hydrological and volcanic processes, with fluvially eroded channels infilled by younger effusive lava flows (Berman & Hartmann, 2002; Burr et al., 2002; Fuller & Head, 2002; Plescia, 2003; Jaeger et al., 2007, 2010; Thomas, 2013; Voigt & Hamilton, 2018). Crater age dating suggests that the youngest lava flows in Elysium Planitia have ages <20 Ma (Berman & Hartmann, 2002; Vaucher et al., 2009; Voigt & Hamilton, 2018) with lava flows in Athabasca Valles perhaps being emplaced as recently as ~2.5 Ma (Vaucher et al., 2009; Jaeger et al., 2010; Golder et al., 2020). Hrad Vallis region has also been interpreted to include phreatomagmatic and mud flow deposits associated with interactions between a dike and ground-ice melted ice to produce a mudflow (Wilson & Mouginis-Mark, 2003b; Morris & Mouginis-Mark, 2006; Pedersen et al., 2010; Pedersen, 2013; Hamilton et al., 2018). Similar flow features have been attributed to volcano–ground-ice interaction in Elysium Planitia (Mouginis-Mark, 1985). Cratered cone groups located throughout Elysium Planitia are interpreted to be pyroclastic deposits generated by explosive lava–water interactions (i.e., volcanic rootless cones; Frey et al., 1979; Lanagan et al., 2001; Keszthelyi et al., 2010; Hamilton et al., 2010, 2011). However, these are secondary landforms and not the products of primary explosive eruptions.

Thus, while the volcanic record of Mars is dominated by effusive volcanism, it also includes a rich record of varied pyroclastic deposits and evidence for magma–water interaction. However, evidence for geologically recent, well-preserved, primary pyroclastic deposits has been lacking. In this study, we present observations of a mantling unit in Elysium Planitia, herein referred to as the Cerberus Fossae mantling unit (CFmu), and test the hypothesis that this may be a geologically recent pyroclastic deposit. The CFmu is a low albedo, high thermal inertia deposit surrounding one of the fissures of the Cerberus Fossae (centered at 165.8° E, 7.9° N), located 25 km west of the young Zunil crater (Figure 1). However, in contrast to the wind streaks observed around craters and other fissures in Elysium Planitia, which all display a dominantly SSW pattern of elongation, the CFmu has an approximately symmetrical distribution around a segment of the Cerberus Fossae and more important, is elongated in the upwind direction. In this study, the statistical analysis of the upwind–downwind elongation patterns of the wind streak features demonstrates that the CFmu is unusual and unlikely to have resulted from ordinary aeolian redistribution of material. Instead, the CFmu appears similar to pyroclastic deposits documented on the Moon and Mercury, which include low-albedo, symmetric features surrounding an obvious source fissure or vent that lacks an associated volcanic construct (Gaddis et al., 1985; Head et al., 2009; Gustafson, 2012). Stratigraphic relationships and absolute age estimates indicate that the CFmu is exceptionally young (53–210 ka) and, if it is a pyroclastic deposit, then it provides evidence of geologically recent explosive volcanism on Mars, suggesting that the Cerberus Fossae and underlying magmatic source may still be active today.

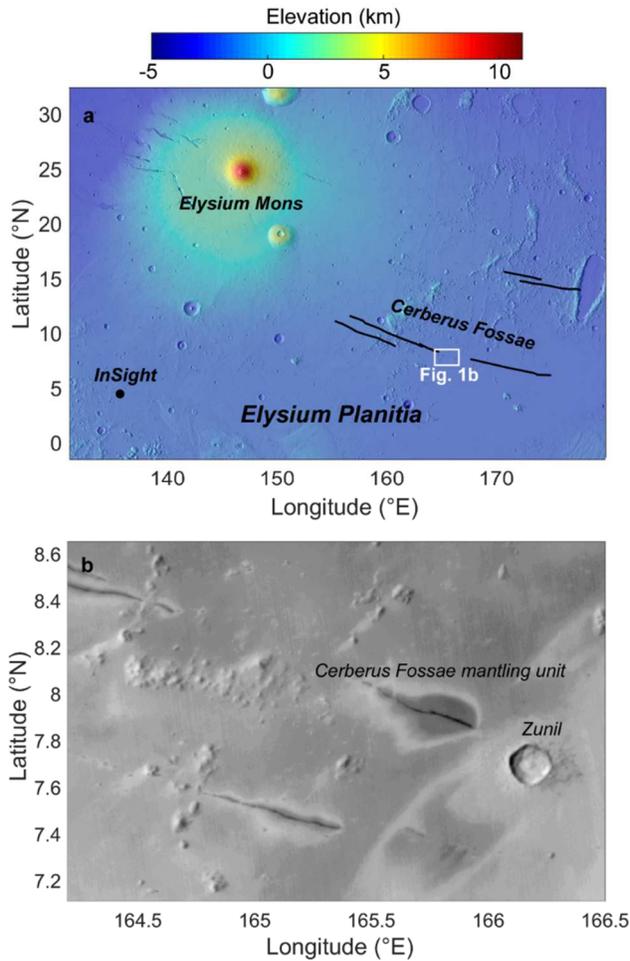

**Figure 1. (a)** The broader regional context and location of the unit and the Cerberus Fossae (black lines) within Elysium Planitia, overlain on a colorized hillshade of Mars Orbiter Laser Altimeter (MOLA) topography. **(b)** Magnified view of a Mars Orbiter Camera (MOC) Wide Angle image mosaic showing the Cerberus Fossae mantling unit, 10 km diameter Zunil crater, and nearby Cerberus Fossae.

## 2. Methodology

### 2.1. Analysis of the morphology and physical properties

*Mars Reconnaissance Orbiter* (MRO) High Resolution Imaging Science Experiment (HiRISE) data (image ID ESP_016519_1880) was used to analyze the surface textures on the mantling unit and map primary craters on the unit, which were used to derive age, thickness and volume estimates for the unit. The characteristics of the impact crater population are critical for constraining both the thickness and age of the mantling unit. The deposit is associated with two populations of small primary impact craters (D > 1 m): bright ejecta craters with higher albedo ejecta blankets, and dark ejecta craters lacking distinctive ejecta blankets. In contrast to the secondary craters from nearby Zunil crater (discussed further in Section 3.5), which are typically elongated along one axis and have irregular outlines, the primary craters mapped on the deposit show no preferential orientation and continuous crater rims. The roll-over in the size-frequency distribution of the primary crater population occurs at ~2 m, indicating that the inclusion of smaller crater does not impact the crater retention modeling age estimates. Furthermore, there is a consistent difference in the lower envelop of the bright ejecta crater and dark ejecta crater diameters, suggesting that the bright and dark ejecta craters are distinct populations. The bright ejecta craters are inferred to post-date the CFmu and to have excavated into underlying brighter materials. The dark ejecta craters are inferred to either pre-date the mantling unit and have been thinly mantled by it, to have not excavated fully through the mantling unit, or have been subsequently modified so as to have indistinct ejecta blankets. Bright ejecta craters are primarily found north of the fissure, likely due to aeolian modification south (downwind) of the fissure. More discussion of these crater populations can be found in Section 3.3.

*Mars Odyssey* Thermal Emission Imaging System (THEMIS) day and nighttime thermal infrared mosaics and MRO Context Camera (CTX) images were used as base maps for mapping inferred isopach geometries, analyzing aeolian features in the surrounding region to derive wind streak distances, and to broadly investigate the morphology of the mantling unit. MRO Compact Reconnaissance Imaging Spectrometer for Mars (CRISM) visible–near infrared (VIS–NIR) hyperspectral data were used to analyze the mineralogy of the unit. The CRISM data (image ID FRT00016467) was corrected for photometric and atmospheric effects using the volcano scan atmospheric correction method (McGuire et al., 2009). Individual spectra were obtained over regions of interest and ratioed with dusty pixels in the corresponding rows (Pelkey et al., 2007), to compare with spectra of other low albedo regions and type spectra from other CRISM observations (Viviano-Beck et al., 2014; Cannon et al., 2017).

**2.2. Isopach and thickness determination**

Thickness estimates were derived using the excavation depths of bright ejecta craters, which provided an upper bound for the unit thickness (see Sections 3.3 and 3.4). We assume a crater excavation depth to transient diameter relationship of 0.1 (Melosh, 2011) and a transient to final apparent crater diameter scaling of 1.1 (Stewart & Valiant, 2006). Although the crater depth-to-diameter relationship may differ for an unconsolidated or loosely-consolidated material, this scaling relationship provides an upper bound for such materials. We determine the thickness as a function of area using the lower diameter envelope of the bright ejecta craters.

The population of bright ejecta craters as a whole shows a relationship of increasing minimum diameter closer to the fissure. Due to the low density of bright ejecta craters on the CFmu and the fact that any one crater only provides a bound on the maximum thickness at the impact site, an *a priori* assumption regarding the spatial variation in the thickness of the deposit must be made to bin craters together for this particular analysis. We assume two different tephra distribution scenarios when creating thickness isopachs for the deposit. First we assume that the asymmetry of the deposit was a product of the distribution of material during the eruption and isopach contours that follow the shape of the CFmu. Second we assume that the distribution of material was symmetric about the source fissure and redistribution of material occurred post-eruption and elliptical isopach contours. We first bin bright ejecta craters based on distance from the fissure in 500 m increments. Using a method similar to that of Fassett et al. (2011), we then find the lower envelope in which only crater diameters that are smaller than the smallest crater from the previous distance bin are included, moving away from the fissure. We then fit each selected crater with an isopach contour for the two dispersal scenarios described above.

We then fit the data to Earth-based empirical relationships between isopach area and deposit thickness for similar eruptions to estimate the volume. As discussed above, the bright ejecta crater population, used to determine the thickness of the deposit, shows a relationship of decreasing thickness away from the fissure and is well fit by the two methods discussed below. Thus, while differences in the gravity and atmospheric conditions between Mars and Earth will influence the distribution of pyroclastic material (e.g., Wilson & Head, 1994; Fagents & Wilson, 1996; Glaze & Baloga, 2002; Wilson & Head, 2007), our analysis suggests that the CFmu is well described by Earth-based empirical relationships of thickness decrease from the source vent. Two methods are used to estimate the thickness and volume of the deposit based on empirical relationships between the isopach area and deposit thickness for eruptions on Earth: the exponential method (Pyle, 1989) and the Weibull method (Bonadonna & Costa, 2012). The exponential method applies an exponential fit to the plot of the ln(thickness)–area$^{1/2}$, then extrapolates to the intercept (eruption source) to determine the maximum thickness. The derived thickness relationship is then:

$$T = T_0 e^{-k\sqrt{A}} \tag{1}$$

where $T_0$ is the maximum thickness determined as the y-intercept. From this fit, we determine a thickness half-distance:

$$b_t = \frac{\ln(2)}{k\sqrt{\pi}} \tag{2}$$

where $k$ is the magnitude of the slope of the ln(thickness)–area$^{1/2}$ plot. Integration over the total area of the deposit ($A$) yields a volume estimate for the deposit:

$$V = \frac{T_0}{\alpha} \int_0^{\sqrt{A}} x^2 e^{-kx} dx \tag{3}$$

where $\alpha$ is related to the isopach eccentricity ($\alpha = a/b$).

A Weibull relationship has been shown to match a range of eruptions on Earth in which the thickness of the tephra deposit is related to the square root of the isopach area by:

$$T = \theta(\sqrt{A}/\lambda)^{n-2} e^{-(\sqrt{A}/\lambda)^n} \tag{4}$$

where $n$ is a shape parameter, $\lambda$ is a characteristic length scale of deposit thinning in km, and $\theta$ is a thickness scale in cm. The best-fit parameters were determined using the methodology of Bonadonna & Costa (2012), minimizing the residual between the observed and calculated thickness values. The volume of a deposit of area $A$ is thus:

$$V = \frac{2\theta\lambda^2}{n}(1 - e^{-(\sqrt{A}/\lambda)^n}) \tag{5}$$

To investigate the sensitivity of the thickness and volume estimates to the assumptions described above, we also derived a best-fit thickness based on the smallest 10% of bright ejecta craters in a particular bin rather than using the thresholding approach of Fassett et al. (2011).

## 2.3. Age determination

We used stratigraphic markers from the secondary craters and rays of Zunil to derive the relative age of the CFmu. Absolute age estimates were based on the population of small (D <100 m) primary craters on the feature. Performing a detailed survey of both bright ejecta and dark ejecta primary craters on the CFmu, we derived age estimates based on the Hartmann production function for small craters (Hartmann, 2005). We performed a survey of craters on the main deposit using HiRISE image ESP_00016519_1880, which covers ~50 km² of the CFmu at a digitizing scale of 1:1000. Crater clusters were corrected by calculating for the effective diameter using the relationship $D_{eff} = (\sum D_i^3)^{1/3}$ (Ivanov et al., 2008, 2009). Modeled crater retention ages were determined only on the area north of the fissure (29 km$^2$) due to the paucity of bright ejecta craters to the south of the fissure. Given that aeolian reworking appears to have darkened the crater ejecta blankets to reduce the population of bright ejecta craters south of the fissure, we cannot rule out the possibility that some dark ejecta craters north of the fissure could be modified bright ejecta craters that post-date the mantling unit. Age estimates were determined for all craters interpreted as primary craters as well as a separate model crater retention age for only the population of bright ejecta craters.

Based on the apparent young age of the CFmu, we also calculated a model crater retention age using a published estimate of the present-day cratering rate (Daubar et al., 2013). Observations of fresh craters (Daubar et al., 2013) suggests that the present-day cratering rate is less than estimates from the other production functions for Mars (Hartmann, 2005), which would increase the age estimates based on the crater population on the deposit. However, more recent work (Daubar et al., 2016) suggests that the rapid fading of the blast zone albedo around recent small craters may explain the discrepancy between these two crater populations (Hartmann, 2005; Daubar et al., 2013) and partly explain the shallower present-day crater size–frequency distribution. Given the uncertainty in the interpretation of the present-day cratering rate and the better fit of the Hartmann (2005) production function to the crater size-frequency distribution, we focus our interpretations of absolute ages on the latter. For completeness, however, we present age estimates using both methods.

## 3. Observations and analysis

### 3.1. Morphology and thermophysical properties

The CFmu exhibits a low albedo (~0.15), high thermal inertia (~200 J m$^{-2}$ K$^{-1}$ s$^{-1/2}$) (Putzig et al., 2005), and smooth surface in comparison to the surrounding plains. The CFmu extends 5.7 km northeast and 12.1 km southwest from a 17.3 km long section of a 34 km long fissure of the Cerberus Fossae (Figure 1b). The CFmu is nearly symmetric about the fissure and is characterized by a clearly defined rounded upwind margin, suggesting that it is sourced from the fissure, unlike the asymmetric aeolian scour and sand streaks extending downwind from the other fossae and topographic landforms in the region (see further discussion on aeolian features in Section 4.1).

The CFmu is surrounded by a higher relative albedo (>0.2), low thermal inertia (<100 J m$^{-2}$ K$^{-1}$ s$^{-1/2}$) halo (Figure 2a, b), similar to typical martian dust (Ruff & Christensen, 2002; Putzig et al., 2005). We interpret the high albedo material as a layer of fine-grained dust underlying and exposed at the edge of the low albedo deposit (Figure 2c). The preservation of indurated dust in the ejecta blankets of Zunil secondary craters outside the deposit supports the presence of a thicker dust layer covering Elysium Planitia in the past (McEwen et al., 2005). General circulation models predict dust deposition in this region at obliquities >35° (Newman et al., 2005), which last occurred prior to ~3 Ma (Laskar et al., 2004), and subsequent deflation at lower obliquities. Although we cannot rule out the possibility that the bright, high albedo material is a finer grained portion of the CFmu, the evidence for a previous regional dust mantle and the excavation of bright material by craters on the deposit (Figure 2b) favor the dust layer interpretation. Regional-scale effusive flow features on the surrounding volcanic plains and some rays of secondary craters associated with nearby craters are buried and muted by the CFmu and (based on the thickness of the CFmu as shown below) also the underlying dust mantle (discussed further in Section 3.3). These observations indicate that the CFmu mantled a former regionally extensive dust layer, preserving it locally beneath the current and former extent of the mantling unit while the dust was removed from the surroundings.

While much of the CFmu and surrounding region are covered in a veneer of dust (Ruff & Christensen, 2002), recent wind activity on the unit appears to have removed this dust immediately southwest of the fissure (Figure 3a). A CRISM spectrum over this exposed portion of the unit is consistent with high-calcium pyroxene (HCP; Figure 3b; Pelkey et al., 2007; Viviano-Beck et al., 2014), although a mixture of HCP and glass cannot be ruled out (Cannon et al., 2017). Nearby low albedo wind streaks and fresh impact crater surfaces do not exhibit a HCP signature (Figure 3c), though an HCP signature has been observed in other freshly exposed volcanic surfaces, as well as young reworked basaltic material such as wind streaks and dunes (Mustard et al., 2005) including the ripples and dunes on the floor of the fissure. While we cannot conclusively rule out the possibility that the observed HCP signature is due to redistribution of basaltic sand from the floor of the fissure, we deem it more likely that the HCP-bearing material is a fresh exposure of the CFmu deposit based on the symmetry of the deposit and the lack of similar spectral signatures in dark streaks extending from other features in the area.

Curvilinear troughs and ridges, 10s of meters in wavelength and approximately perpendicular to the fissure, are observed throughout the CFmu (Figure 4b), but are concentrated near the fissure. Based on the estimated thickness of the CFmu (see discussion below), we suggest that this pattern originates in the underlying dust layer or the basal volcanic surface. The morphology, geometry, and scale of this curvilinear texture could be consistent with thinly mantled inflated lava flows (Bleacher et al., 2017; Voigt & Hamilton, 2018), fine-scale secondary spatter around fissures on Earth (Jones et al., 2018), transverse aeolian ridges (TARs) in the underlying unconsolidated dust layer (Balme et al., 2008; Zimbleman, 2010; Kerber & Head, 2012; Geissler, 2014), or may be associated with the deposit itself.

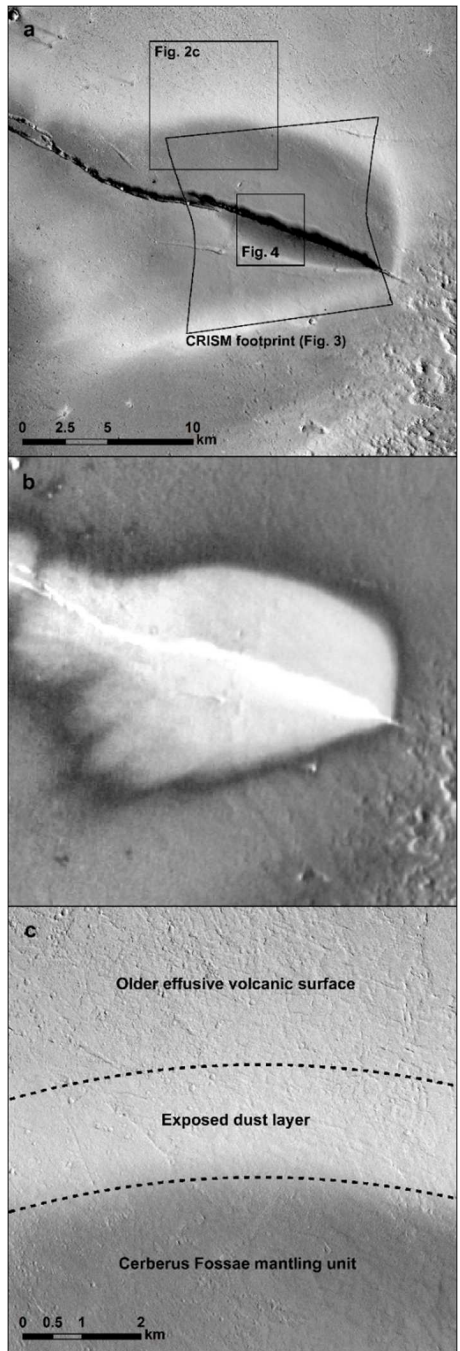

**Figure 2. (a)** Context Camera (CTX) image (B18_016519_1879_XI_07N194W) and **(b)** Thermal Emission Imaging System (THEMIS) daytime infrared of the unit showing the symmetric nature of the unit around a Cerberus Fossae. **(c)** High Resolution Imaging Experiment (HiRISE) imagery (ESP_016519_1880) showing that the appearance of the Cerberus Fossae mantling unit differs from that of the surrounding volcanic plains, mantling both older dust and lava layers.

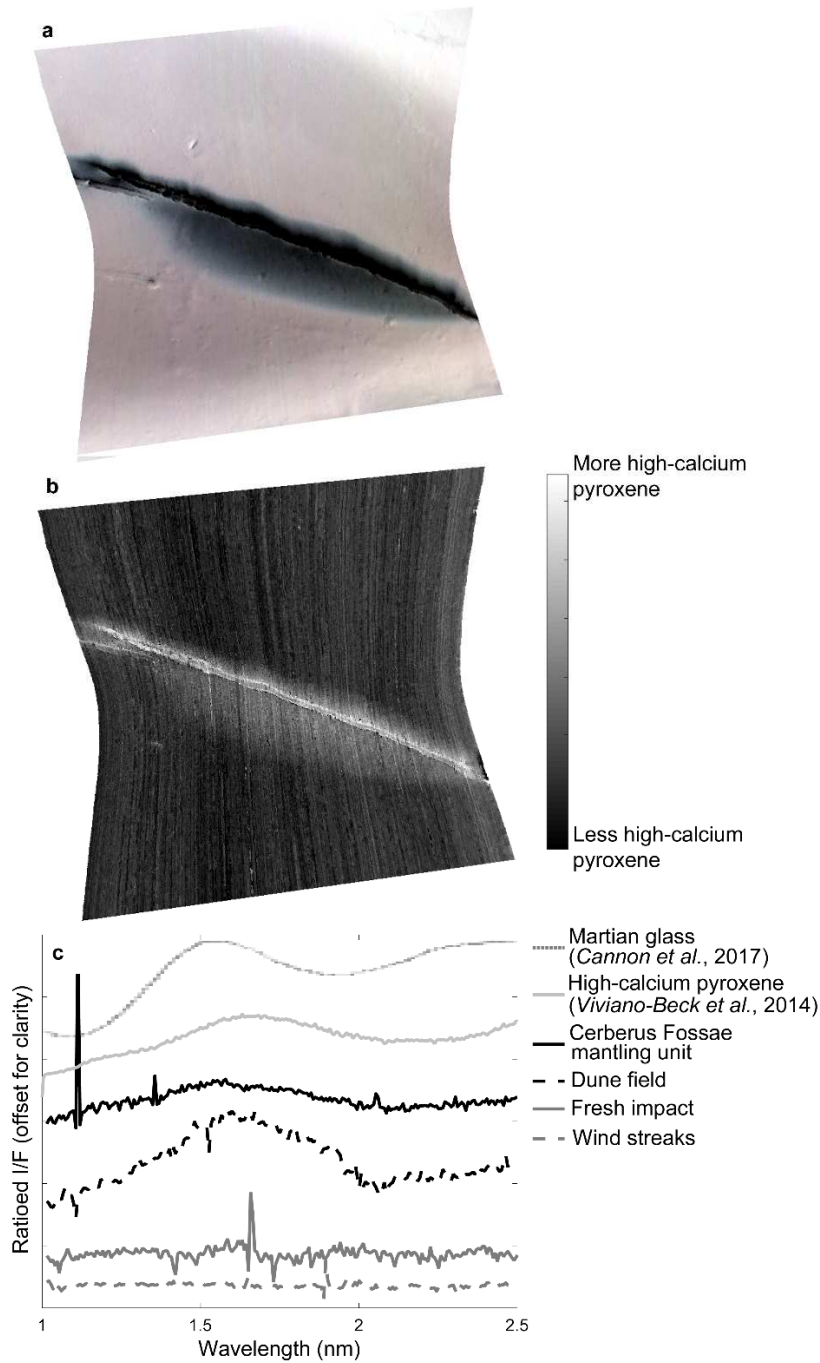

**Figure 3. (a)** Compact Reconnaissance Imaging Spectrometer for Mars (CRISM) image (FRT00016467) over the unit with 2.5950, 1.5066, and 1.0800 nm bands in red, green, and blue respectively and **(b)** high-calcium pyroxene (HCP) map highlighting the low albedo exposed region near the fissure. **(c)** Individual ratioed spectra are shown for the Cerberus Fossae mantling unit, a nearby dune field, a fresh impact, dark dust scour wind streaks, and a dark volcanic surface. These spectra are compared with CRISM type spectrum for high-calcium pyroxene (Pelkey et al., 2007) and martian glass (Cannon et al., 2017).

## 3.2. Aeolian modification

Recent wind activity was observed on this deposit, resulting in a visible darkening of the area immediately north and south of the fossa between CTX images P13_006221_1881_XN_08N194W and B18_016519_1879_XI_07N194W, which were acquired in 2007 and 2009, respectively. Darkening was observed ~1 km from the fissure in the upwind direction and ~2 km in the downwind direction (Figures 4a). We attribute this darkening to scour of a thin mantle of dust that was deposited during the dust storm of 2007, which brightened the surface of the Cerberus Plains and the CFmu.

Textures consistent with aeolian ridges are observed throughout the deposit. These small features are ~3–5 m in length and ~1–2 m in width, with crests roughly oriented NNE–SSW (Figure 4c, black arrows). These ridges tend to form in linear chains from the southeast to the northwest, and are similar to TARs observed elsewhere on Mars (Balme et al., 2008; Zimbelman, 2010; Geissler, 2014). In addition, sets of linear features orthogonal to the aeolian bedforms oriented from the southeast to the northwest are interpreted as forming through aeolian erosion or redistribution of material (Figure 4c, white arrows). Although the visible and thermal properties of the CFmu are consistent with minor induration of the deposit or coarser grained tephra, these aeolian features indicates some minor reworking and mobility of the deposit. Thus, we infer that the deposit must contain an appreciable fraction of sand-sized grains and/or be only weakly indurated. These features appear to be the youngest textures on the deposit.

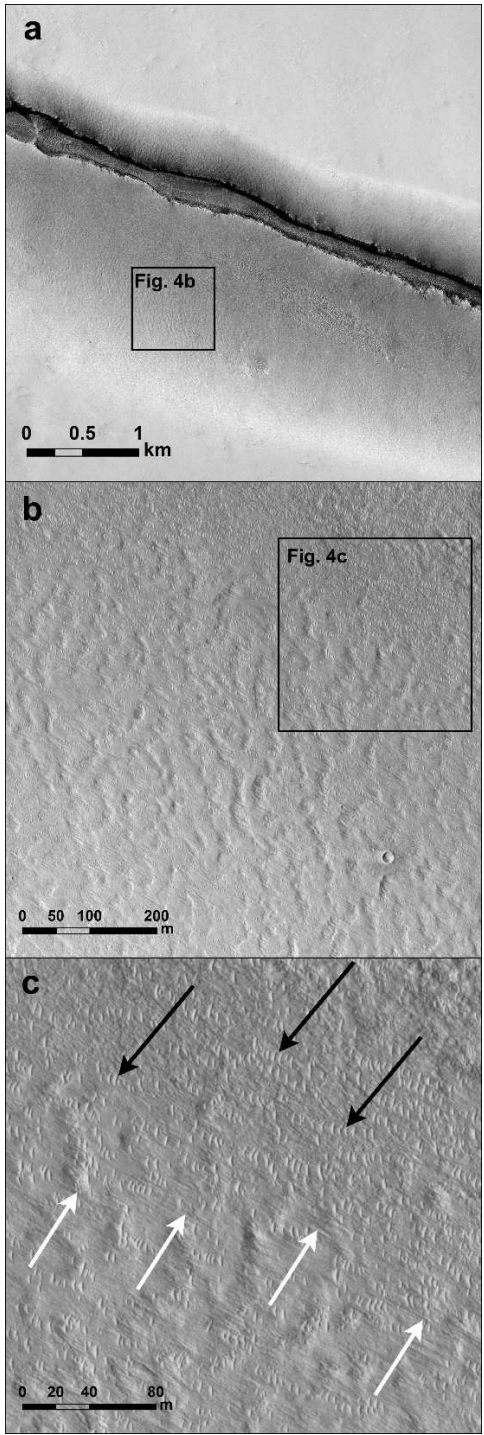

**Figure 4.** **(a)** Source fissure of the CFmu and darkening of the surface proximal to the fissure either from scour of a thin dust mantle or transport of basaltic sand from the fissure floor (note that the CFmu as a whole extends far beyond the edges of the image). **(b)** Curvilinear texture on the unit (165.877° E, 7.882° N), **(c)** small aeolian ridges (trending NNE–SSW) and linear erosional features (trending NW–SE). Examples of the small aeolian ridges and linear erosional features are noted by the black and white arrows respectively.

### 3.3. Primary craters on the mantling unit

As previously discussed in Section 2.1, two distinct primary crater populations are observed within the CFmu: bright ejecta craters with high albedo ejecta blankets and dark ejecta craters that lack distinct ejecta blankets and have no albedo contrast relative to the surrounding unit (Figure 5). There are three possible interpretations for impacts responsible for the dark ejecta craters. The impacts could have pre-dated the deposition of the CFmu and the craters are mantled by the thin CFmu; the impacts could have post-dated the deposition of the CFmu, but were too small to excavate to the underlying dust layer; or the impacts could have post-dated the deposition of the CFmu and excavated to the underlying dust layer, with the bright ejecta being reworked or obscured by subsequent aeolian processes. The smallest dark ejecta craters may not have excavated through the deposit and thus may pre- or post-date the deposit. The larger dark ejecta craters are consistent with the interpretation that these pre-date the deposition of the CFmu and are mantled by it, as they lack obvious ejecta blankets (Figure 5a, b). However, bright ejecta around some of the craters that appears modified by aeolian processes but not yet erased is consistent with aeolian reworking/obscuring and possible darkening the bright ejecta over time (Figure 5c). These observations suggest that dark ejecta craters both pre- and post-date the mantling unit.

While there is some uncertainty in the interpretation of the dark ejecta craters, bright ejecta craters on the CFmu have clearly excavated through the mantling deposit to an underlying dust layer, preserved as a higher albedo ejecta blankets (Figure 5c, d). Bright ejecta craters are primarily concentrated to the north of the fissure (Figure 6) and appear to be randomly distributed, though there is some spatial dependence on the location and size of the bright ejecta crater population.

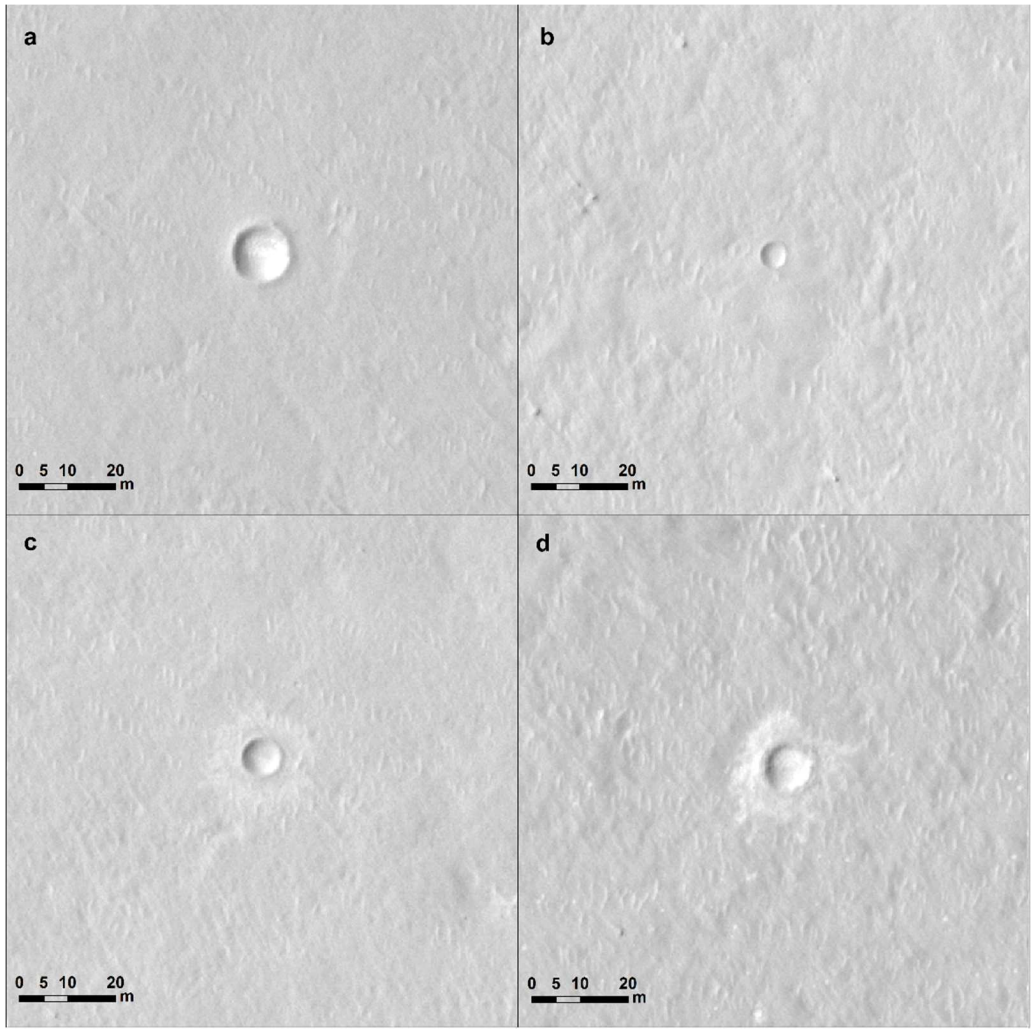

**Figure 5. (a)–(b)** Craters that lack distinct ejecta relative to the surrounding Cerberus Fossae mantling unit (referred to as dark ejecta craters) either pre-date the deposition of the unit and are thinly mantled by the deposit or have been reworked by aeolian processes. **(c)** Aeolian reworking appears to be modifying some bright ejecta, consistent with the observed bright and dark crater distribution. **(d)** Bright ejecta craters are interpreted to have excavated through the deposit to an underlying dust layer. All images are from HiRISE image ESP_016519_1880.

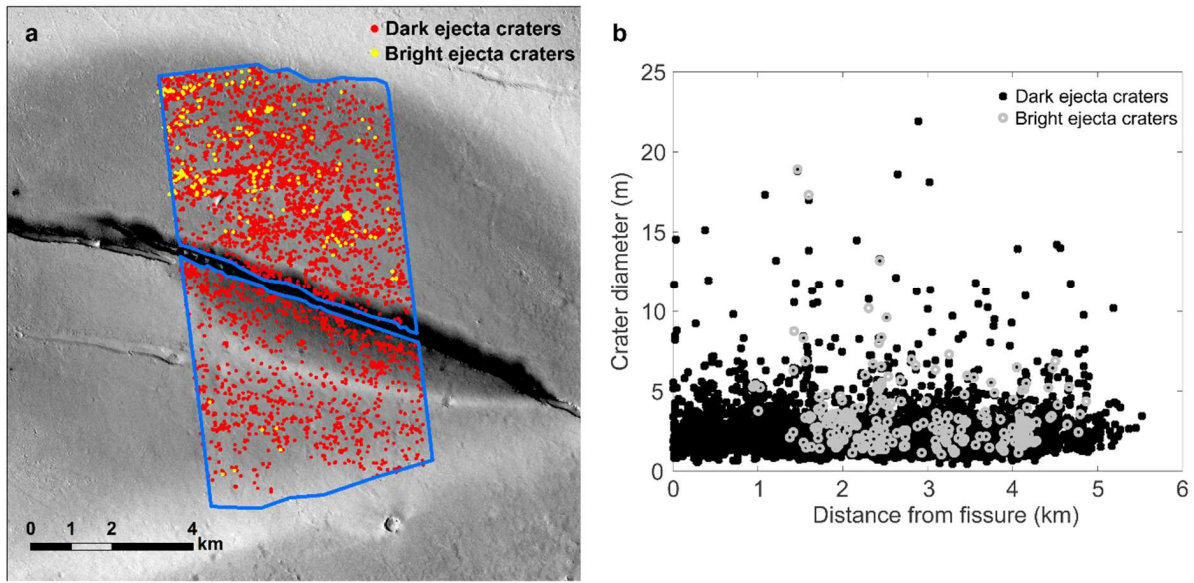

**Figure 6.** Thickness and age analyses were conducted using **(a)–(b)** the distribution of bright ejecta craters and craters that lack obvious ejecta (referred to as dark ejecta) on the unit.

To the south bright ejecta craters are less abundant (particularly at diameters <3 m) and absent altogether within 3 km of the fissure (Figure 6a). This may be consistent with a thicker deposit in the south, though there is not a distinct rollover at larger dark ejecta crater diameters compared to the north as would be expected if the deposit was thicker in the south. We conclude that the modification of crater ejecta blankets by aeolian processes downwind of the fissure is the most likely explanation for the discrepancy in the distribution and sizes of bright and dark ejecta craters between the areas north and south of the fissure. This conclusion is consistent with the observed aeolian modification (see Section 4.1) downwind of the Cerberus Fossae and other topographic obstacles in Elysium Planitia. For this reason, age and thickness estimates are based on the northern portion of the deposit.

Alternatively, bright ejecta crater could preserve a remnant dust layer preserved in the ejecta blanket similar to the Zunil secondaries in this region. However, a similar bright ejecta crater population located outside of the deposit is not observed, the minimum diameter of the bright ejecta craters is systematically greater than the dark ejecta craters, the bright ejecta craters diameters increase approaching the fissure, and the CFmu clearly mantles a brighter material and the surrounding lava surface. This indicates that the bright ejecta craters are exposing an underlying dust layer and can be used to determine an upper bound on the thickness of the CFmu.

### 3.4. Thickness estimates

We constrained the thickness of the mantling unit using the bright ejecta craters inferred to have excavated through the unit into the underlying dust layer. The diameter of the smallest bright ejecta craters increases closer to the fissure, and bright ejecta craters are not observed within 1 km of the fissure (Figure 6b), supporting a thickening of the unit with proximity to the fissure (though an increase in deposit strength and resistance to erosion closer to the fissure cannot be ruled out). The excavation depths of the bright ejecta craters to the north of the fissure (Figure 6b) provide an upper bound estimate for the mantling unit thickness and indicate that the majority of the unit has a maximum thickness of 0.1 to 0.4 m and is consistent with an exponential decay in thickness with distance from the fissure.

### 3.5. Age estimates

Stratigraphic relationships provide the best constraint on the relative age of this unit. The CFmu is stratigraphically above both the surrounding volcanic plains, and the now largely removed regional dust deposit. The bright raised ejecta of the Zunil secondaries, due to armoring of a former regional dust deposit by the secondary ejecta (McEwen et al., 2005), allow this population to easily be delineated from other crater populations (Figure 7a) and provides an important regional stratigraphic marker with an estimated age of ~0.1–1 Ma (Hartmann et al., 2010; Williams et al., 2014). Rayed Zunil secondaries typically have irregular outlines and are elongated along one axis (Figure 7b). While Zunil secondaries are observed on the volcanic plains surrounding the CFmu (Figure 2a, b), Zunil secondaries are not as easily identified on the CFmu. Heavily mantled secondary craters in the deposit appear to have some preferential orientation (Figure 7c), consistent with being secondaries of either Zunil or the more distant (~1500 km) and older Corinto crater, the latter of which also has an extensive field of secondary craters (Watters et al., 2017). Elongated linear features 100s of meters in width with a rubbly texture, consistent with the morphologies of rays emanating from Zunil, cross the unit and have no albedo contrast relative to the unit (Figure 7d). The diameters of secondary craters within these rays exceed the diameters of bright ejecta primary craters, and thus excavation and exposure of the underlying dust would be expected if the rays post-date the deposit. Rather, we interpret these Zunil rays as predating the deposit but post-dating the underlying dust layer, thus being only thinly mantled by the deposit.

Given the possibility that some dark ejecta craters may post-date the deposit but have been modified by aeolian reworking, we prefer the interpretation that the age of the pyroclastic deposit lies somewhere between the model ages obtained for the bright ejecta craters and all craters. We calculate a lower bound on the age for the northern portion of the CFmu of 53 ± 7 ka using only bright ejecta craters north of the fissure and an upper bound on the age of 210 ± 12 ka by including both dark and bright ejecta craters north of the fissure, both using the Hartmann (2005) production function (Figure 8a). For the present-day cratering rate (Daubar et al., 2013) we find a model age for the northern portion of the deposit of 780 ± 120 ka using only bright ejecta craters north of the fissure and 3.3 ± 0.6 Ma using both dark and bright ejecta craters north of the fissure (Figure 8b), though this chronology is not preferred for reasons discussed previously. The crater size frequency distribution for both crater populations are better fit by the slope of the Hartmann (2005) production function than by the present-day production function. We also note that the crater size frequency distribution for the CFmu is statistically indistinguishable from crater counts on the ejecta blanket and floor of Zunil crater (Hartmann et al., 2010). Given the range of age estimates of Zunil crater (~0.1–1.0 Ma), the age estimates derived for the CFmu and uncertainties associated with the absolute age estimates, the absolute ages of the CFmu and Zunil crater are indistinguishable.

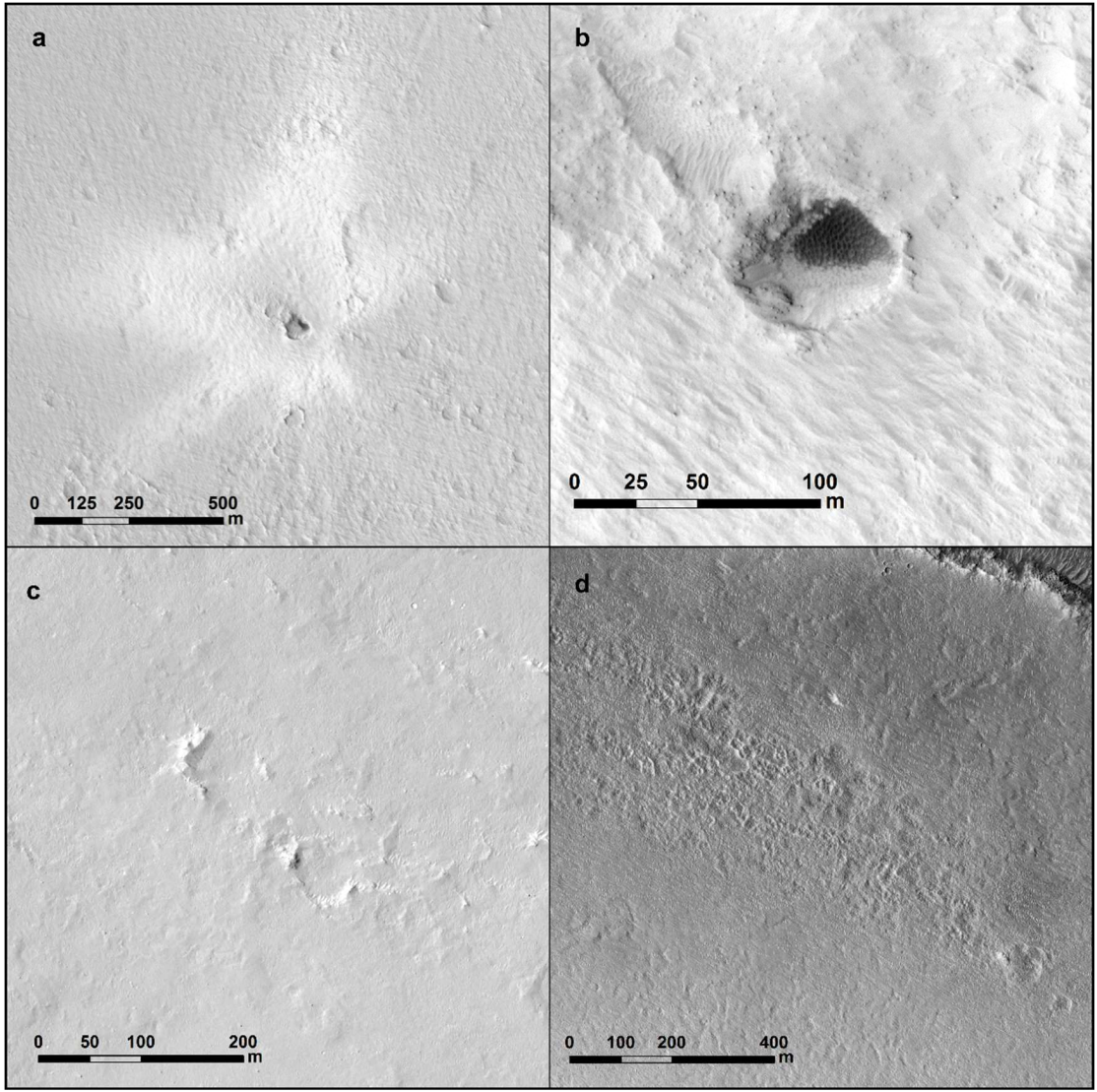

**Figure 7.** Examples of Zunil secondary craters outside (to the northwest) of the Cerberus Fossae mantling unit (CFmu; HiRISE image PSP_006221_1880), showing **(a)** the bright raised ejecta halo found around many Zunil secondaries (165.693° E, 8.181° N) and **(b)** a secondary crater elongated along one axis with an irregular rim (165.679°E, 8.081°N). **(c)** A potential secondary crater of Zunil within the deposit that has been subsequently buried (165.867° E, 7.94° N), although its orientation is also consistent with the nearby Corinto crater. **(d)** Ray-like features in the unit that orient back to Zunil (165.904° E, 7.879° N) and are morphologically consistent with rays observed in the Zunil near-field. The ray-like texture appears to superimpose the surrounding curvilinear texture but does not excavate through or otherwise disturb the thin mantling unit. Panels **(c)** and **(d)** are from HiRISE image ESP_016519_1880.

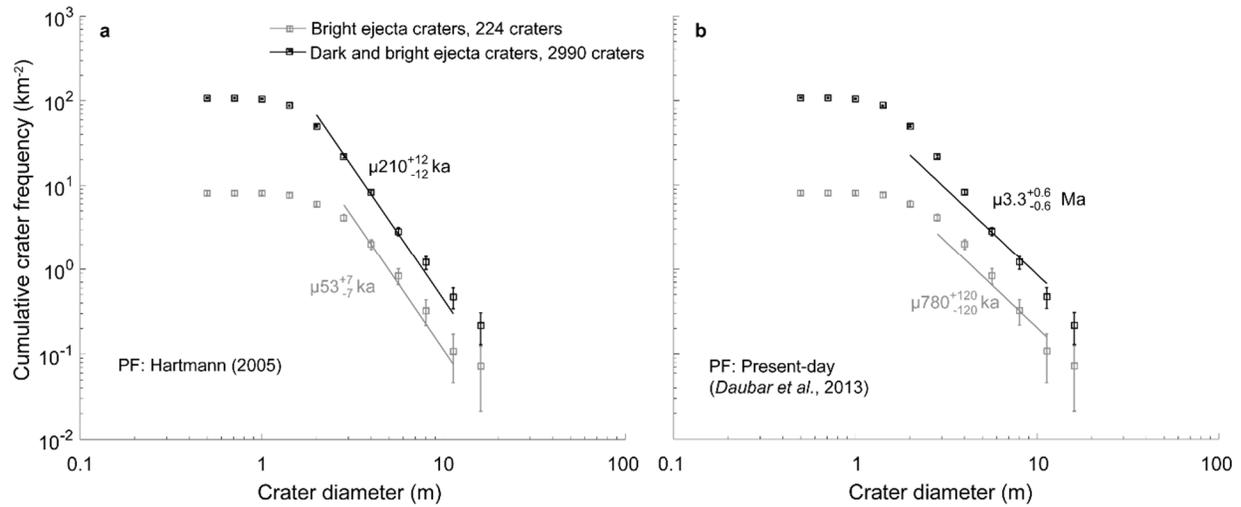

**Figure 8.** Modeled crater retention ages using **(a)** the Hartmann, (2005) production function and **(b)** the present-day production function (Daubar et al., 2013) for bright ejecta and combined bright and dark ejecta crater populations. Errors are formal crater counting errors and do not reflect larger uncertainties on small crater (<100 m) counting.

Caution is required when interpreting these small-crater and small-area crater retention ages. The stated uncertainties are the formal uncertainties resulting from counting statistics and do not account for other variables such as atmospheric breakup of small meteoroids, secondary contamination and degradation of small craters, which can result in large errors in absolute age estimates. Crater counting on small areas (<100 km$^2$) in particular can result in a wide range of age estimates (Warner et al., 2015). Uncertainties on the shape of the production function size distribution for craters less than 100 m, can result in a factor of 10 uncertainty on modeled age estimates (Hartmann, 2005), though the good match between the observed and model production functions supports the resulting ages. Perhaps the largest source of uncertainty is the possibility that dark ejecta craters also post-date the features and have lost their distinctive bright ejecta blankets. Taking both bright and dark ejecta craters into consideration, the age of the deposit is likely in the range of 46–222 kyr. While the absolute ages of this and other young surfaces are highly uncertain (Daubar et al., 2014; Williams et al., 2014), the stratigraphic positions relative to the Zunil crater clearly demonstrate this feature to be exceedingly young.

## 4. Interpretation

The analyses above clearly show that the CFmu is a young dark deposit mantling the underlying dust layer and volcanic surface. We now consider two possible interpretations for the origin of the CFmu: an aeolian origin due to either deposition or erosion of sand-sized material and a volcanic origin from a geologically young pyroclastic eruption. In the aeolian scenario, the low albedo mantling deposit is the result of either wind driven redistribution of basaltic sand sourced from the fissure or wind driven scour of the dusty plains. Alternatively, in the volcanic scenario, the deposit is the result of explosive volcanic activity, which erupted tephra from a segment of the Cerberus Fossae, emplacing it onto the surrounding terrain. While Elysium Planitia contains numerous examples of recent aeolian modification, we show below that the morphology, thermophysical properties, and geometry are inconsistent with an aeolian origin. Thus, we favor the interpretation that this feature is a fissure-fed pyroclastic deposit.

### 4.1. Comparison with aeolian features in Elysium Planitia

Topographically-influenced aeolian scour and deposition features initiating at discrete topographic obstacles are found throughout Elysium Planitia, near the CFmu. Wind streaks associated with craters, fissures, and positive relief in Elysium Planitia predominantly trend northeast to southwest, though smaller streaks occur at several angles oblique to this trend (primarily around craters less than 100 m in diameter), indicating multiple prevailing wind directions (Figure 9b). Low-albedo crater-related wind streaks in Elysium Planitia are either due to dust deflation or the transport and deposition of crater or fissure sourced basaltic sands (Greeley et al., 1978; Rafkin et al., 2001). Dust deflation wind streaks around different fossae or other obstacles superficially resemble the CFmu, with a low albedo zone and a surrounding bright albedo halo from the redistribution of the thin regional dust cover. Darkening of the surface due to the transport of basaltic sand generally forms long linear features extending from sand deposits within fissures and craters. These features tend to initiate from nick points in crater rims or from discrete points along fissures where irregularities in the walls (i.e., kinks and ramps) or the ends of the fissures allow for easy transport of sand material to the surrounding plains (Figure 9c–k). The Cerberus Fossae wind streak features of both types are variably elongated in the southwest (downwind) direction (~1–25 km) with a limited and relatively uniform elongation in the northeast (upwind) direction (~1 km; Figure 10). This is consistent with aeolian scour and/or deposition altering the surface far downwind of topographic obstacles, while limited upwind scour and/or deposition may be controlled by eddies generated by topographic relief. In contrast, the CFmu is more symmetric being elongated ~6 km in the upwind direction and ~12 km in the downwind direction (Figure 9a). The CFmu is more strongly symmetric when measured relative to the direction perpendicular to the fossae, extending 5.7 km to the NNE (maximum distance) and 6.6 km to the SSW, with the latter being a more diffuse boundary. This elongation of the deposit in the upwind direction is not compatible with a purely aeolian origin. The upwind elongation of the CFmu is 20 standard deviations greater than the upwind elongation of the wind streak population, which has a mean upwind elongation distance of 0.4 km and standard deviation of 0.27 km. We perform a Grubb's outlier test (Grubbs, 1969) on the CFmu upwind elongation, confirming that the CFmu is not part of the Cerberus Fossae wind streak population at the 99% confidence level. Either the CFmu was formed by a unique wind regime that affects only this particular fissure and not those to the east or west, or it is not an aeolian feature. Furthermore, the CFmu tapers out at the end of the fissure where transport of basaltic sand out of the fissure should be easiest and instead appears to be sourced from the fissure as a whole. The modest asymmetry of the CFmu is consistent with either limited post-depositional aeolian redistribution of the deposit or the influence of wind on the deposition of the pyroclastic material at the time of the eruption. The two fossae immediately to the southwest of the CFmu exhibit similar but smaller albedo patterns (Figure 1, 9c–d). These features are similarly symmetric about their respective fissures, but do not extend as far in the upwind directions. Although these may be similar fissure-sourced pyroclastic deposits, a purely aeolian origin cannot be ruled out. If these features are pyroclastic deposits, they must be exceedingly thin and not underlain by a thick dust deposit as they do not show any evidence of mantling the surficial lava flow textures.

### 4.2. Pyroclastic origin for the Cerberus Fossae mantling unit

The approximately symmetric distribution of the CFmu around one of the Cerberus Fossae is morphologically consistent with a fissure-fed tephra deposit and is located in a region in which such deposits might be expected to form based on abundant evidence for recent volcanic and hydrologic activity (e.g., Berman & Hartmann, 2002; Burr et al., 2002; Fuller & Head, 2002; Plescia, 2003; Thomas, 2013; Voigt & Hamilton, 2018). The areal extent of the CFmu (221 km$^2$) is comparable to that of small to medium sized pyroclastic deposits on the Moon (Gaddis et al., 2003) (mean of 1721 km$^2$, with a range of 3 to 49,013 km$^2$) and somewhat smaller than that of deposits on Mercury (Kerber et al., 2011b) (mean of 2670 km$^2$, with a range of 317 to 19,466 km$^2$). The lateral extent of the deposit (~5–7 km) is consistent with modeled radial plume expansion (~5–8 km) up to the maximum height of convective entrainment of ~10 km on Mars (Glaze & Baloga, 2002), though deposition may have included vent-proximal dispersal of ballistic pyroclasts as well (Wilson & Head, 2007). The thickness constraints for the deposit are well-fit by both exponential and Weibull models of pyroclastic deposits (Figure 11). The thickness–area$^{1/2}$ relationships obtained for the deposit provides volume estimates of $1.7 \times 10^7$, $1.7 \times 10^7$, and $1.1 \times 10^7$ m$^3$ for the exponential method and $2.2 \times 10^7$, $2.0 \times 10^7$, and $2.8 \times 10^7$ m$^3$ for the Weibull method, assuming irregular isopachs, elliptical isopachs, and elliptical isopachs using the smallest 10% of bright ejecta craters in each bin, respectively (Figure 11b, c, d). These volumes are on the order of $10^7$ m$^3$, regardless of the methodology used to determine the isopach areas. Volume estimates for the deposit are comparable to those for Hawaiian and Subplinian volcanic eruptions of basaltic magma on Earth of $10^5$ to $10^8$ m$^3$ (Houghton et al., 2013). The thickness profile of the deposit, derived from the thickness–area$^{1/2}$ relationship and quantified as the thickness half-distance ($b_t$ = 1.9 km), is consistent with Plinian and Subplinian eruptions on Earth (Pyle, 1989).

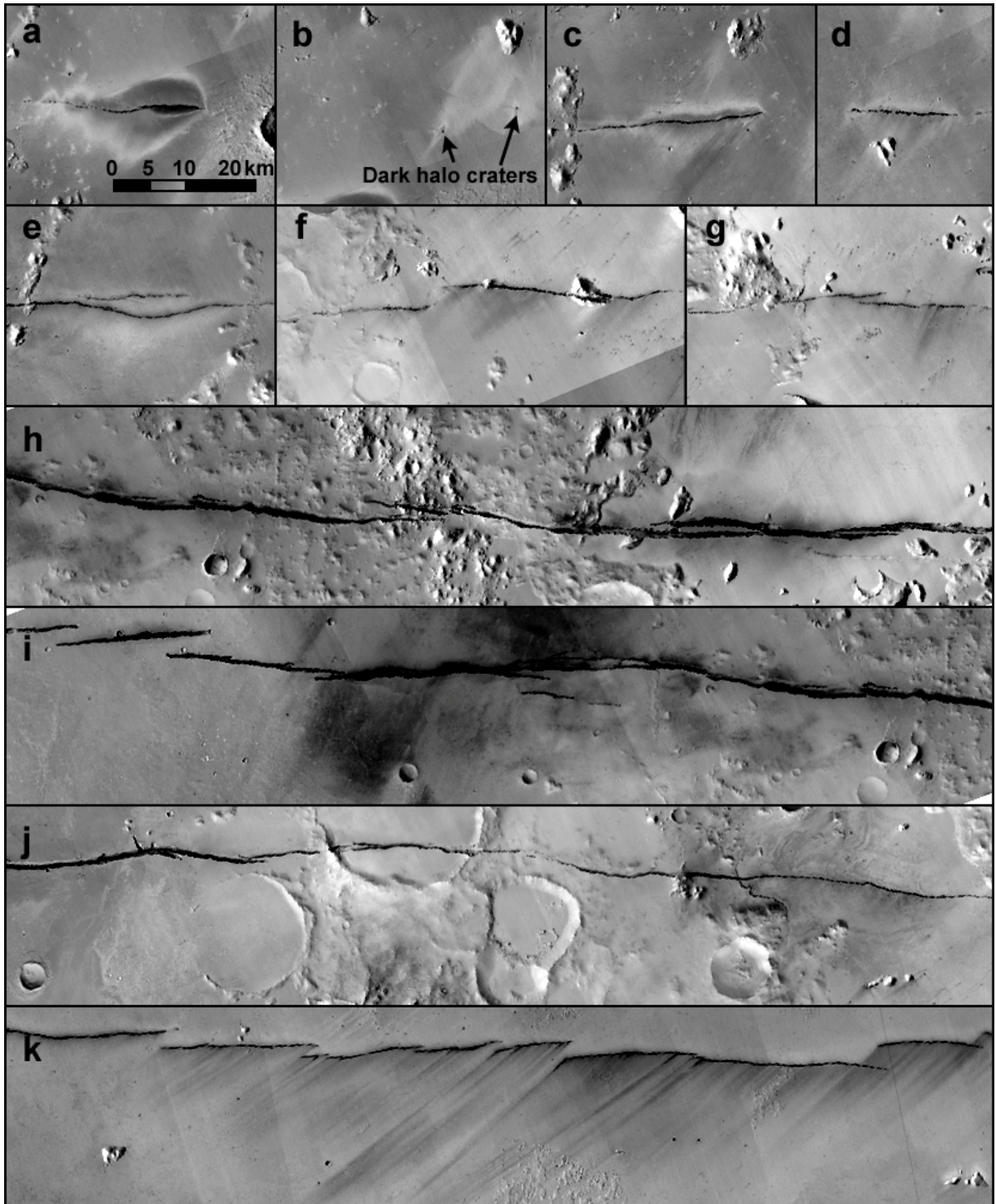

**Figure 9.** CTX mosaics of **(a)** the Cerberus Fossae mantling unit (CFmu), **(b)** dark halo craters to the north of the CFmu (the largest two noted in the image), and **(c)**–**(k)** other major fissure windstreaks in Elysium Planitia. Only two short fissure segments closest to the CFmu, **(c)** – **(d)**, display some elements of the CFmu morphology, but the upwind dimension of these deposits does not fundamentally differ from the aeolian modification in Elysium Planitia. All images are at the same scale. CTX mosaic credit: NASA/JPL/MSSS/The Murray Lab.

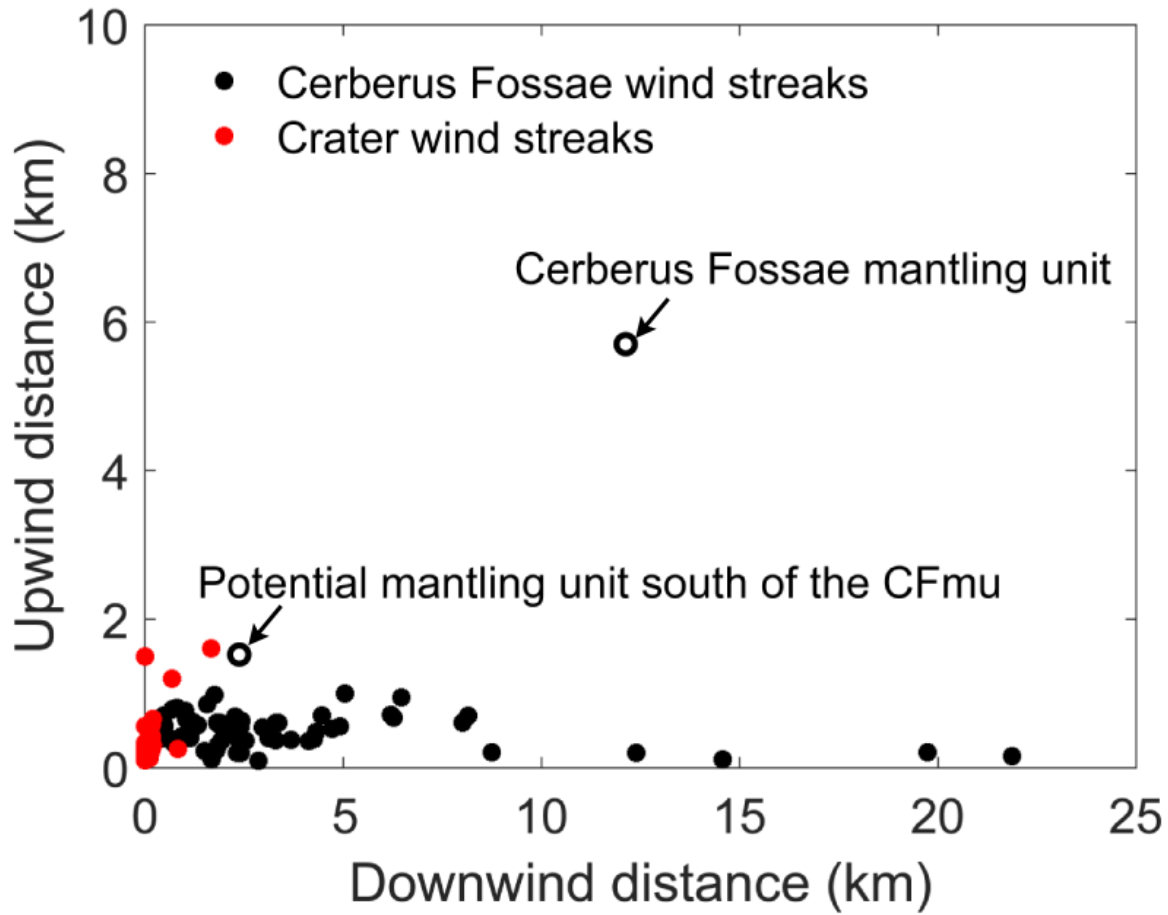

**Figure 10.** The relationship between the upwind (NNE) and downwind (SSW) streak lengths relative to the obstacle at the source of the streak indicates that the Cerberus Fossae mantling unit is unrelated to aeolian processes responsible for the low albedo observed around many of the Cerberus Fossae.

Furthermore, exponential thinning is consistent with volcanic fall deposits on Earth. Alternatively, multiple linear segments could also fit the ln(thickness)–area$^{1/2}$ plots (Figure 11b, c, d), which could indicate ballistic dispersal of pyroclasts near the source and fall deposition in the more distal regions (Bonadonna & Houghton, 2005; Biass et al., 2019). Differences in gravity and atmospheric conditions between Mars and Earth will influence processes of fragmentation, tephra dispersal, and pyroclastic deposition (Wilson & Head, 1994; Fagents & Wilson, 1996; Greeley et al., 2000; Glaze & Baloga, 2002; Wilson & Head, 2007), but overall the characteristics of pyroclastic fall deposits on Earth and Mars should be fundamentally similar (Wilson & Head, 1983; Wilson & Head, 2007).

Visibly, the unit has diffuse margins and appears to mantle the underlying dust and volcanic plains (Figure 2a, c), which is consistent with pyroclastic deposits on Earth, the Moon, and Mercury (Gaddis et al., 1985; Head et al., 2009; Kerber et al., 2011b). While there is no volcanic construct associated with the CFmu (Figure 4a), pyroclastic deposits on the Moon and Mercury often form around a source vents with no detectable volcanic construct (e.g., Gustafson et al., 2012; Head et al., 2009). The lack of a volcanic construct is consistent with the characteristics of some deposits associated with phreatic and phreatomagmatic eruption phases on the Earth, which mantle the existing topography beneath thin tephra deposits without constructing a high-standing vent-proximal edifice (e.g., Mattsson & Hoskuldsson, 2011; Hughes et al., 2018; Zawacki et al. 2019). The low albedo and high thermal inertia are consistent with a coarse-grained or indurated fine-grained basaltic material, while the limited aeolian mobility of the deposit indicates a substantial component of ash-sized material. The thermal inertia corresponds to an effective grain size of ~100 µm (Presley & Christensen, 1997); however, given the likely effects of dust on the surface and the possibility of an indurated fine-grained deposit, it is possible the deposit contains a mixture of coarser and finer grain particles. Grain size variability with distance from the source fissure could not be determined. The presence of HCP where the surface of the unit has been exposed is similar to what is observed for young pyroclastic deposits on the Moon (Gaddis et al., 2003). If the HCP was sourced by the eruption, it may originate from either the primary magma or ejected country rock material from the existing volcanic plains.

Given the observations of a pyroxene-rich, dark mantling deposit distributed quasi-symmetrically around and thinning away from a volcanic fissure in a system known to have sourced some of the youngest eruptions on Mars, the simplest explanation for the origin of the deposit is that it is a thin pyroclastic deposit formed during an explosive volcanic eruption. Although a lower atmospheric pressure on Mars relative to the Earth, is expected to favor the development of explosive eruptions driven by juvenile magma volatiles (Wilson & Head, 1983), it is likely that external water may have played a role in the explosivity of this eruption through the interaction of magma with an ice-rich regolith (Moitra et al., 2019). Elysium Planitia includes widespread evidence for the release of subsurface-sourced water (Berman & Hartmann, 2002; Burr et al., 2002; Fuller & Head, 2002; Manga, 2004; Voigt & Hamilton, 2018) as well as evidence for deep and shallow ground ice (McEwen et al., 2005; Keszthelyi et al., 2010). Therefore, as an alternative to a purely magmatic eruption, it is possible that meltwater generated in association with the eruption may have interacted with the ascending magma to enhance fragmentation via phreatic or phreatomagmatic processes.

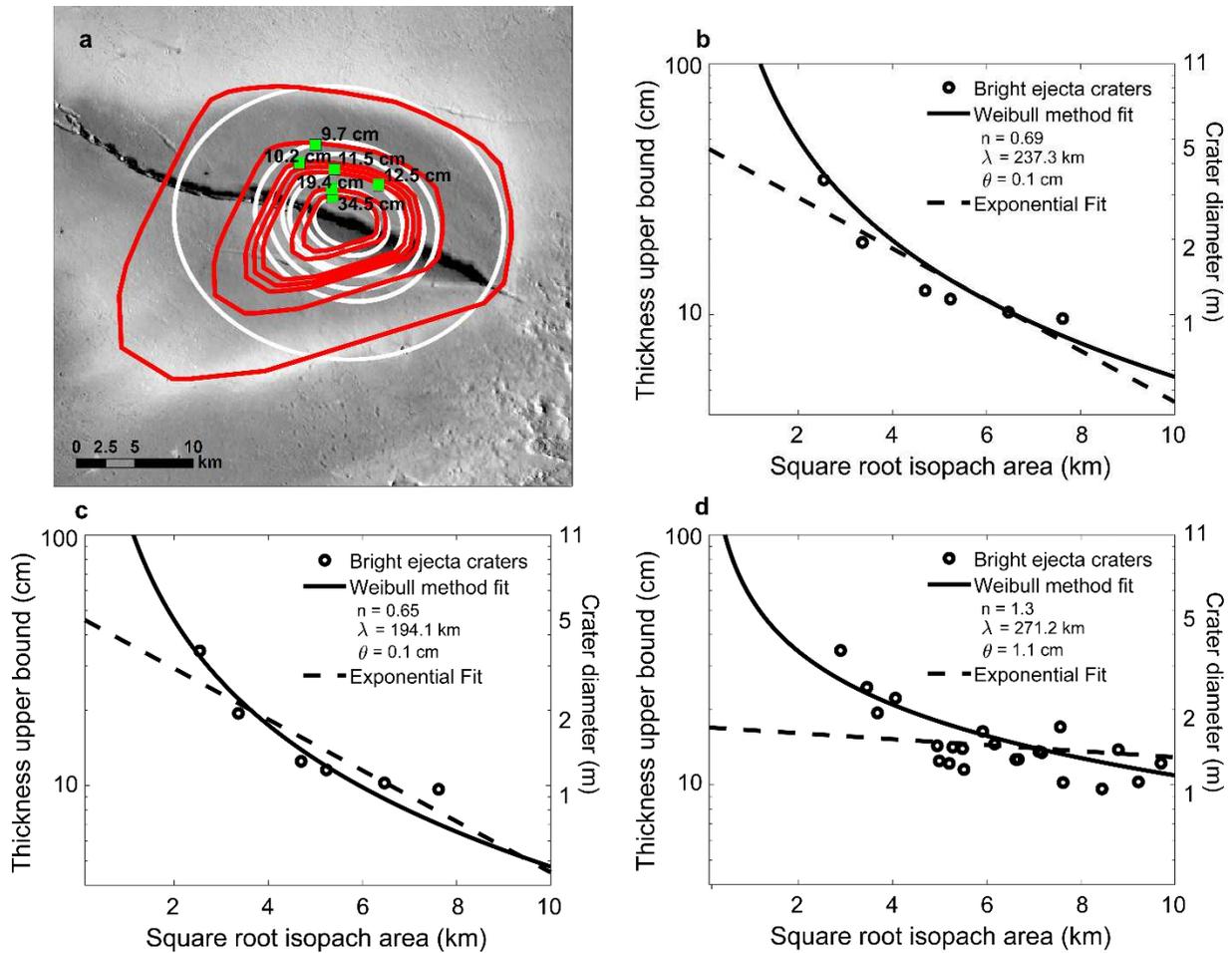

**Figure 11. (a)** Isopach maps for the elliptical (white) and Cerberus Fossae mantling unit (CFmu; red) geometries determined from the bright ejecta craters (green points) observed on the CFmu. Values shown are the corresponding thickness estimates derived from the bright ejecta craters. The thick outer lines represent the total area of the deposit for each isopach shape. Based on these isopachs, best-fit thickness estimates are derived for using the Weibull (Bonadonna & Costa, 2012) and exponential methods (Pyle, 1989) using **(b)** a geometry matching the shape of the CFmu **(c)** elliptical isopachs and **(d)** the smallest 10% of bright ejecta craters. Best-fit parameter values for the Weibull method are provided.

## 5. Conclusions and implications

The Cerberus Fossae mantling unit is interpreted to be the youngest volcanic product discovered on Mars to date. The deposit is morphologically and thermophysically consistent with a late Amazonian fissure-fed pyroclastic deposit in Elysium Planitia. While pristine pyroclastic deposits are well documented on the Earth, Moon, and Mercury, no such deposit of comparable preservation state has been documented on Mars until now. Previous work interpreted thin, low-albedo features around inferred pits and fissures in Elysium Planitia to be young pyroclastic deposits (Roberts et al., 2007), but higher resolution images show these to be thin mantles of basaltic sand redistributed from deposits within secondary craters of Zunil (Figure 7a, b). Although explosive eruptions may occur more frequently than observations would suggest, the resulting thin pyroclastic deposits could be easily obscured by erosion (Golombek et al., 2014), volcanic resurfacing (Morgan et al., 2013; Voigt & Hamilton, 2018), or mantling by dust (Ruff & Christensen, 2002; Newman et al., 2005) and may be lost from the geologic record. Thus, the CFmu may simply be preserved owing to its extremely young age.

Geologically recent near-surface magmatic activity in Elysium Planitia, combined with evidence for recent groundwater-sourced floods (Burr et al., 2002; Head et al., 2003), which may have been triggered by dike intrusions (Hanna & Phillips, 2006), raises important implications regarding the subsurface habitability on Mars. Dike-induced melting of ground ice and hydrothermal circulation could generate favorable conditions for recent or even extant habitable environments in the subsurface. These environments would be analogous to locations on Earth where volcanic activity occurs in glacial environments such as Iceland, where chemotrophic and psychrophilic (i.e., cryophilic) bacteria thrive (Cousins & Crawford, 2011). Subsurface microbial communities found in basaltic lavas on Earth (McKinley et al., 2000) are also aided by hydrothermal circulation of groundwater through porous basalt (Storrie-Lombardi et al., 2009; Cousins & Crawford, 2011). Recent or ongoing magmatic activity on Mars could also provide a source of transient methane releases to the atmosphere (Formisano et al., 2004; Fonti & Marzo, 2010) through direct volcanic outgassing or, more likely, serpentinization reactions (Atreya et al., 2007).

Given the young age of the deposit, it is possible that the deeper magma source that fed the deposit could still be active today and could generate seismicity observable by the Seismic Experiment for Interior Structure (SEIS) instrument on the *Interior Exploration using Seismic Investigations, Geodesy, and Heat Transport* (InSight) lander (Lognonné et al., 2019). Seismicity related to magma transport and chamber pressurization has been linked to active volcanism on Earth (e.g., Battaglia et al., 2005; Grandin et al., 2012; Carrier et al., 2015). Magma-induced seismicity along rift zones can result in small to moderate earthquake magnitudes ($M_w < 6$). Dike-induced faulting and seismicity (Rubin & Gillard, 1998; Taylor et al., 2013) associated with this young magmatic activity is also possible.

The theoretical detection threshold for the SEIS instrument is $M_w$ ~3 at 40° epicentral distance (Lognonné et al., 2019). Based on the location of the InSight lander relative to the CF mantling unit (~1750 km to the southwest or ~31° epicentral distance), small to intermediate Marsquakes ($M_w < 6$) and potentially microseismic events ($M_w < 2$) associated with magmatic activity may be observed by the SEIS instrument. Thus far, the only two locatable Marsquakes detected by the SEIS instrument on InSight were sourced from the Cerberus Fossae region (Giardini et al., 2020), supporting the possibility that this region remains magmatically active today.

**Acknowledgements.** This work was supported by grant NNX17AL51G to JCAH from the Mars Data Analysis Program. HiRISE, CTX, CRISM and THEMIS data used for this work are publicly available through the NASA Planetary Data System Mars Node.

Wilson, L., Mouginis-Mark, P. J., Tyson, S., Mackown, J., & Garbeil, H. (2009). Fissure eruptions in Tharsis, Mars: Implications for eruption conditions and magma sources. *Journal of Volcanology and Geothermal Research*, *185*(1–2), 28–46. https://doi.org/10.1016/j.jvolgeores.2009.03.006

Zawacki, E. E., Clarke, A. B., Arrowsmith, J. R., Bonadonna, C., & Lynch, D. J. (2019). Tecolote volcano, Pinacate volcanic field (Sonora, Mexico): A case of highly explosive basaltic volcanism and shifting eruptive styles. *Journal of Volcanology and Geothermal Research, 379*, 23-44. http://doi.org/10.1016/j.volgeores.2019.04.011

Zimbelman, J. R. (2010). Transverse Aeolian Ridges on Mars: First results from HiRISE images. *Geomorphology*, *121*(1–2), 22–29. https://doi.org/10.1016/j.geomorph.2009.05.012